\documentclass[conference]{IEEEtran}
\IEEEoverridecommandlockouts
\usepackage{cite}
\usepackage{amsmath,amssymb,amsfonts}
\usepackage{algorithmic}
\usepackage{graphicx}
\usepackage{textcomp}
\usepackage{xcolor}
\usepackage[a4paper, total={184mm,239mm}]{geometry}

\usepackage{booktabs}
\usepackage{tikz}
\usepackage{pgfplots}
\pgfplotsset{compat=1.18} 
\usepackage{url}

\def\BibTeX{{\rm B\kern-.05em{\sc i\kern-.025em b}\kern-.08em
    T\kern-.1667em\lower.7ex\hbox{E}\kern-.125emX}}
\begin{document}

\title{SPOILER-GUARD: Gating Latency Effects of Memory Accesses through Randomized Dependency Prediction\\
\vspace{-18pt}

}

\author{
\IEEEauthorblockN{Gayathri Subramanian\thanks{\vspace{-0.35cm}Gayathri Subramanian is the corresponding author.}, Girinath P, Nitya Ranganathan, Kamakoti Veezhinathan, Gopalakrishnan Srinivasan}
\IEEEauthorblockA{Department of Computer Science and Engineering, Indian Institute of Technology Madras\\
\{gayathri, cs22s021, kama, sgopal\}@cse.iitm.ac.in,
nitya\_rise@icsrpis.iitm.ac.in
}
\vspace{-35pt}
}

\maketitle

\begin{abstract} 

Modern microprocessors depend on speculative execution, creating vulnerabilities that enable transient execution attacks. Prior defenses target speculative data leakage but overlook false dependencies from partial address aliasing, where repeated squash and reissue events increase the load–store latency, which is exploited by the SPOILER attack. We present SPOILER-GUARD, a hardware defense that obfuscates speculative dependency resolution by dynamically randomizing the physical address bits used for load–store comparisons and tagging store entries to prevent latency-amplifying misspeculations. Implemented in gem5 and evaluated with SPEC 2017, SPOILER-GUARD reduces misspeculation to 0.0004 percent and improves integer and floating-point performance by 2.12 and 2.87 percent. HDL synthesis with Synopsys Design Compiler at 14 nm node demonstrates minimal overheads - 69 ps latency in critical path, 0.064 square millimeter in area, and 5.863 mW in power.
\end{abstract}


\vspace{-0.7cm}

\section{Introduction}

\vspace{-0.15cm}

Modern out-of-order microprocessors use speculative execution to hide latency by predicting branch outcomes and memory dependencies. Misspeculation rolls back architectural state, but microarchitectural state such as cache contents remain altered and are typically exploited by transient execution attacks \cite{b1,b2,b3}. Spectre \cite{b4} leverages branch misprediction while Meltdown \cite{b5} exploits deferred exception handling. Speculation also affects memory operations, where loads may bypass preceding stores if memory-dependence predictors within the pipeline do not predict the dependency \cite{b6,b7,b8}. 

SPOILER \cite{b9} exploits false dependencies induced via partially aliased accesses, revealing fine-grained physical address (PA) information. This accelerates virtual-to-physical reverse-engineering by 256$\times$ and eviction-set construction by 4096$\times$, amplifying attacks such as Prime+Probe \cite{b10} and Rowhammer \cite{b11}. Partial address checks in the Memory Order Buffer \cite{b12,b13,b14} can trigger repeated squash-and-reissue cycles, and the partial physical address bits stored in the Store Address Buffer (SAB) make aliasing-induced latency observable.

In this work, we assume an unprivileged attacker on the same physical core, who is capable of allocating memory, crafting load–store sequences, and performing fine-grained timing via \texttt{rdtsc} instruction \cite{b15}. Our key observation is that SPOILER succeeds because the dependence predictor performs static physical address comparisons, producing repeatable squash-and-reissue cycles. Existing defenses \cite{b16,b17,b18,b19,b20,b21} target speculative data leakage by blocking or sanitizing unsafe speculative loads. However, mechanisms like NDA \cite{b22} and SSBD \cite{b23} incur high overhead, up to 125\% in strict modes and 6.6–10.7\% in practical ones. SPOILER-ALERT \cite{b24} embeds a cuckoo filter in the Store Buffer to track store addresses and detect load–store aliasing with 99.99\% accuracy, but cannot block leakage. Intel guidelines \cite{b25} ignore structural leakage whereas speculative-interference studies \cite{b26,b27} fail to address contention effects, leaving SPOILER exploitable. \textbf{SPOILER-GUARD} obfuscates dependency resolution by randomizing the physical address bits used for speculative checks and tagging store entries to prevent repeated replays. The key contributions of our work are:
\vspace{-2pt}
\begin{itemize}
\item SPOILER-GUARD breaks the correlation between partial address aliasing and microarchitectural timing by dynamically randomizing address bits and tagging misspeculated stores to avoid further misspeculation.
\item We implement and evaluate SPOILER-GUARD in gem5 using SPOILER binaries and SPEC 2017 benchmark suite to demonstrate security and performance.
\item SPOILER-GUARD lowers misspeculation to 0.0004\% and improves integer and floating-point performance by 2.12\% and 2.87\% respectively, with negligible area (0.064 mm$^{2}$), power (5.863 mW), and latency (69 ps) overhead.
\end{itemize}

\vspace{-0.35cm}

\section{Proposed Design}
\vspace{-0.5cm}

\begin{figure}[htbp]
    \centering
    \vspace{-0.05cm}
    \includegraphics[width=1.0\linewidth]{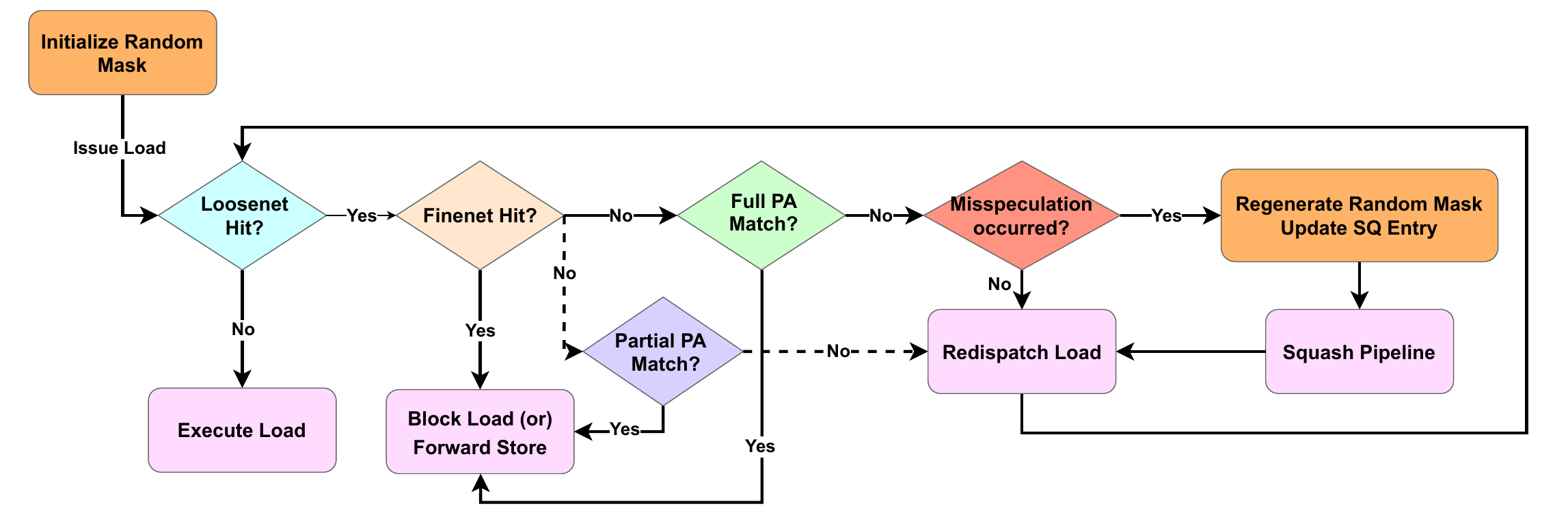} 
    \vspace{-0.85cm}
    \caption{Overview of the proposed SPOILER-GUARD defense.}
    \label{fig:defense_design}
\end{figure}

\vspace{-0.35cm}

In SPOILER, the attacker fills the Store Buffer with aliased stores and issues probe loads with 1 MB aliasing, causing repeated squash-and-reissue cycles. SPOILER-GUARD breaks this correlation by randomizing the partial address bits used for dependency resolution. The fixed 8-bit comparison is replaced with comparison of dynamically selected 12-bits (determined by the \textit{mask}), reducing the attacker’s probability of inferring the aliasing pattern from $1/256$ to $1/4096$. In simulation, the mask is generated using a Mersenne Twister Pseudo Random Number Generator (or PRNG) seeded with high-entropy operating system sources \cite{b28}. In hardware, equivalent randomness can be realized using a True Random Number Generator (TRNG) with entropy conditioning and expansion through a cryptographically secure PRNG \cite{b29}.

SPOILER-GUARD adds three microarchitectural changes: (1) SAB entries gain a PC tag and spoiler-vulnerability flag, (2) partial physical address field is widened to 12 bits, and (3) LSQ integrates mask-generation logic with a remasking controller. An initial mask is set at system setup, with remasking triggered on misspeculation. When a load is executed and encounters a loosenet hit but a finenet miss, the dependency predictor flags a potential dependency and performs a masked 12-bit partial physical address comparison. A partial hit triggers speculative store-to-load forwarding, and the corresponding SQ entry is tagged with the load’s PC. After resolution of the full physical address, the predictor detects misspeculation, sets the vulnerability bit, triggers remasking, and squashes the load, thus preventing repeated aliasing and making any residual latency indistinguishable from normal speculation (see Fig. \ref{fig:defense_design}).

\vspace{-0.3cm}

\section{Evaluation}

\vspace{-0.25cm}

\subsection{Setup and Evaluation}

\vspace{-0.2cm}

We use the gem5 architectural simulator \cite{b30} to model a single-core and single-threaded microarchitecture (downsizing Intel i7-8700), allowing predictable speculative dependency behavior and detailed observation of LSQ interactions. All simulations are run in a full-system Ubuntu Linux x86 environment, capturing realistic system-level interactions to evaluate defense effectiveness. We compare three configurations: M1 - baseline with virtual address (VA) forwarding, M2 - baseline with SPOILER vulnerability, and M3 - SPOILER-GUARD.

\vspace{-0.32cm}

\subsection{Security Analysis}

\begin{figure}
    \centering
    \vspace{-0.5cm}
    \includegraphics[width=1.0\linewidth]{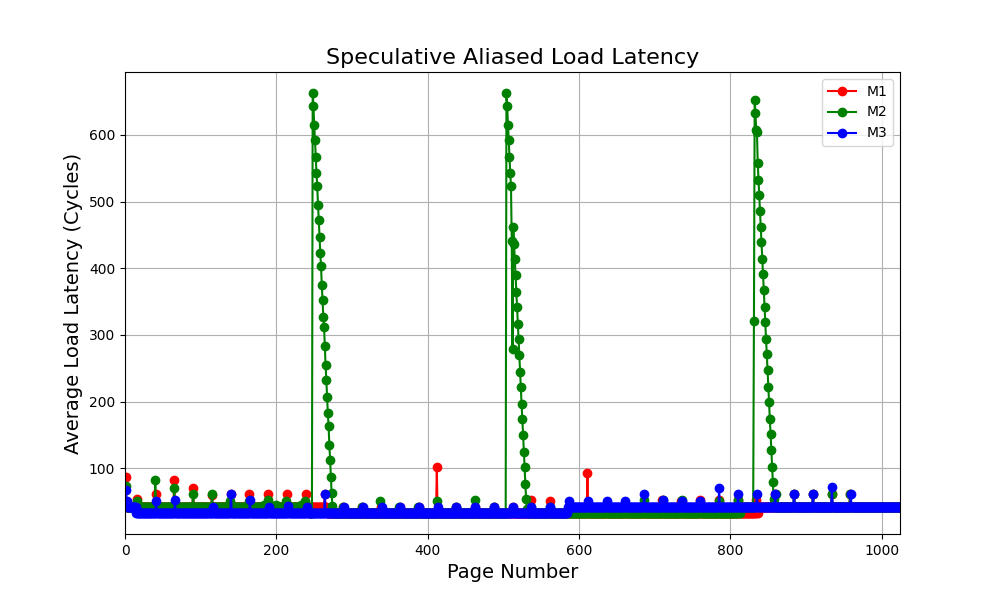}
    \vspace{-0.85cm}
    \caption{Average latency of the speculatively issued malicious load for M1, M2, and M3 configurations.}
    \vspace{-0.7cm}
    \label{fig:latency}
\end{figure}
\vspace{-0.1cm}

The three models are evaluated by their obfuscation of partially aliased load latency and their frequency of false positives from speculative forwarding. A standard attack binary, based on \cite{b9}, allocates a 4KB buffer across 1024 pages, filling the store buffer to induce 4KB and 1MB aliasing, and measuring average latency over 100 rounds per page. This reliably reproduced the attack on Intel i7-8700 and newer processors, including the i9 systems, revealing higher-order physical address bits.

In M1, speculative load latency is uniform with minimal 4KB aliasing delays. In M2, 1 MB aliasing triggers repeated squash-and-reissue (about 255,941 SPOILER-violations and 255,924 attacker-induced stalls), resulting in the high observable latency for these loads. In M3, SPOILER-violations fall to 14 and attacker-induced stalls to 1, with latency of target load matching M1 aside from the unavoidable 4 KB aliasing (see Fig. \ref{fig:latency}). 

In M3, the attacker’s aliasing probability is $P_{M3}(alias)=2^{-12}$ per trial, and remasking on each misspeculation makes attempts statistically independent, collapsing SPOILER’s required temporal correlation. This neutralizes partial physical-address aliasing effects, making speculation indistinguishable from purely VA-based behavior, thereby eliminating the leakage channel and making the defense secure as well as practical.

\vspace{-0.16cm}
\subsection{Performance Analysis}
\vspace{-0.1cm}

\begin{figure}[htbp]
    \centering
    \vspace{-0.4cm}
    \includegraphics[width=1.0\linewidth]{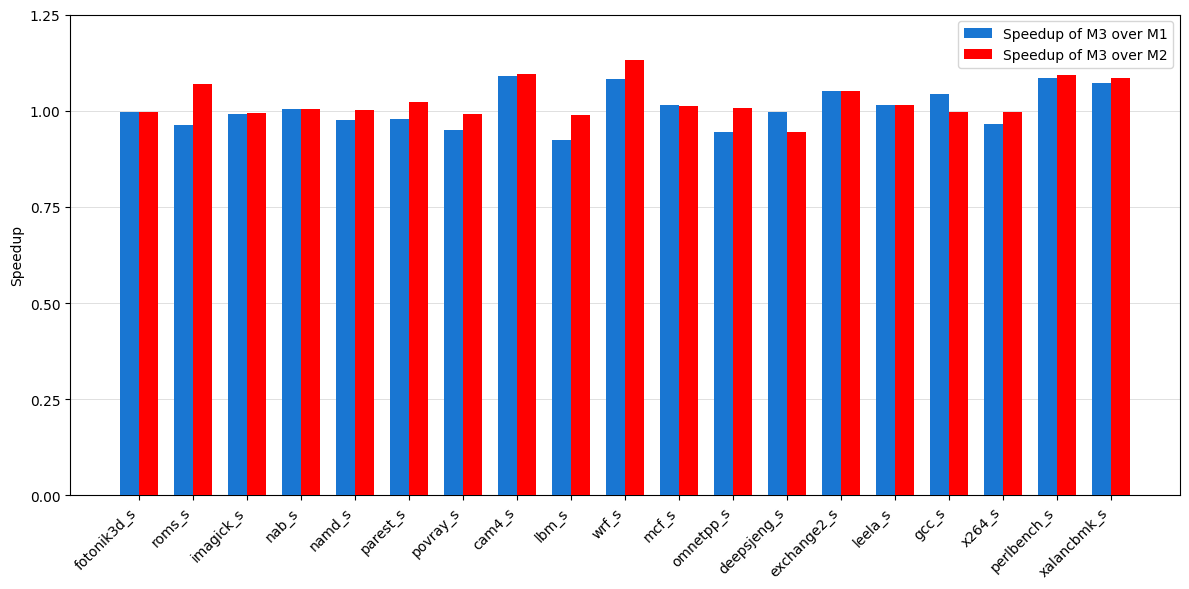}
    \vspace{-0.8cm}
    \caption{Speedup achieved by SPOILER-GUARD for SPEC2017 integer and floating-point benchmark suite.}
    \vspace{-0.55cm}
    \label{fig:combinedspec}
\end{figure}

We evaluated performance using the SPEC CPU2017 suite (compiled using GCC 7.4.0), simulating an average of one billion committed instructions per benchmark. Performance was measured using cycles-per-instruction (CPI), with speedup computed relative to baseline models M1 and M2. Memory and compute intensive workloads were analyzed separately to assess differential responsiveness. SPOILER-GUARD shows overall speedups of 2.87\% for floating-point workloads and 2.12\% for integer workloads. Memory-bound floating-point workloads gain 3.23\%, while memory-bound integer workloads gain 1.13\%. The performance of compute-bound benchmarks remained near baseline. Non-classified workloads benefit more substantially, achieving 7.03\% improvement for floating-point and 3.46\% for integer benchmarks (see Fig. \ref{fig:combinedspec}).

\vspace{-0.05cm}
\subsection{Storage Overhead}
\vspace{-0.1cm}
The enhanced SAB expands the partial physical address field from 8 to 12 bits and introduces two metadata components: a 48-bit load PC tracker and a 1-bit spoiler-vulnerability flag. This adds 53 bits per entry, or 2968 bits ($\approx$0.36\,KB) for a 56-entry SAB. Relative to a typical 32 KB L1 data cache, the SAB incurs less than 0.2\% area overhead while providing significant microarchitectural security enhancements.

\vspace{-0.2cm}

\subsection{Power, Area, and Timing Overhead}
\vspace{-0.1cm}
To evaluate SPOILER-GUARD’s hardware impact, the partial address comparison and SAB remasking logic were implemented in RTL and synthesized using the Synopsys Design Compiler (DC) with a 14 nm standard-cell library. Compared to M2, M3 incurs minimal overheads of 0.064 mm$^{2}$ (less than 0.8\% of a Skylake core’s 8.73 mm² footprint synthesized in a comparable 14 nm process \cite{b31}), 5.863 mW in power, and 69 ps along the critical path (negligible impact), demonstrating that SPOILER-GUARD adds minimal hardware cost while securing speculative dependency resolution.

\vspace{-0.2cm}

\section{Conclusion and future work}
\vspace{-0.1cm}
 We proposed SPOILER-GUARD, a hardware defense that mitigates SPOILER by obfuscating partial address aliasing in LSQ. Applicable to all affected Intel Core microarchitectures, it provides robust security and speedups for memory-intensive workloads. Logic synthesis confirmed minimal area, power, and timing overheads, thereby demonstrating an efficient and practical mitigation. Future work may extend dynamic dependency resolution and store tagging to other speculative attacks.

\end{document}